\documentclass[runningheads]{llncs}
\usepackage[T1]{fontenc}
\usepackage{graphicx}
\usepackage{amsmath}

\usepackage{booktabs}
\usepackage{adjustbox}
\usepackage{hyperref}

\usepackage{listings}
\usepackage{xcolor}
\usepackage{subfig}

\begin{document}
\title{Physics-Aware Compression of Plasma Distribution Functions with GPU-Accelerated Gaussian~Mixture~Models}
\titlerunning{Physics-Aware Compression of Plasma Distribution Functions with GMM}
% If the paper title is too long for the running head, you can set
% an abbreviated paper title here
%
\author{Andong~Hu\inst{1} \and
Luca~Pennati\inst{1} \and Ivy Peng \inst{1} \and
Stefano~Markidis\inst{1}}
\authorrunning{A. Hu et al.}
% First names are abbreviated in the running head.
% If there are more than two authors, 'et al.' is used.
%
\institute{KTH Royal Institute of Technology, Stockholm, Sweden}
\maketitle              % typeset the header of the contribution
\begin{abstract}
Data compression is a critical technology for large-scale plasma simulations. Storing complete particle information requires Terabyte-scale data storage, and analysis requires ad-hoc scalable post-processing tools. We propose a physics-aware in-situ compression method using Gaussian Mixture Models (GMMs) to approximate electron and ion velocity distribution functions with a number of Gaussian components. This GMM-based method allows us to capture plasma features such as mean velocity and temperature, and it enables us to identify heating processes and generate beams. We first construct a histogram to reduce computational overhead and apply GPU-accelerated, in-situ GMM fitting within \texttt{iPIC3D}, a large-scale implicit Particle-in-Cell simulator, ensuring real-time compression. The compressed representation is stored using the \texttt{ADIOS 2} library, thus optimizing the I/O process. The GPU and histogramming implementation provides a significant speed-up with respect to GMM on particles (both in time and required memory at run-time), enabling real-time compression. Compared to algorithms like SZ, MGARD, and BLOSC2, our GMM-based method has a physics-based approach, retaining the physical interpretation of plasma phenomena such as beam formation, acceleration, and heating mechanisms. Our GMM algorithm achieves a compression ratio of up to $10^4$, requiring a processing time comparable to, or even lower than, standard compression engines.

\end{abstract}
\section{Introduction}
Plasma simulations are essential computational tools for understanding the dynamics of space, astrophysical systems, and fusion devices like tokamaks. Among the various computational techniques, the Particle-in-Cell (PIC) method~\cite{hockney2021computer,dawson1983particle} is one of the most advanced and widely used plasma simulation tools for studying the evolution of electrons and protons (called ions in plasma physics) under the influence of electromagnetic fields. The distribution function evolves by sampling computational particles and following their trajectories. The key information from PIC simulations is the data about particle positions and velocities, as it enables us to identify acceleration and heating mechanisms in the plasmas. Nonetheless, the storage requirements and post-processing costs significantly limit the volume of simulation data that can be saved and analyzed. In this work, we develop a compression method based on the Gaussian Mixture Model (GMM)~\cite{bishop2006pattern} to tackle this challenge. Rather than saving all particle information, we perform a GMM operation, obtaining the weight, center, and covariance for a small number of Gaussians used to represent all the data, thus saving the storage by several factors.

In particular, we develop a compression technique to compress particle distribution functions. Electron and ion distribution functions are related to the probability of finding an electron or an ion at a certain position in the phase-space, namely a position (x,y,z) in Cartesian coordinates with certain velocity components (u,v,w). 
In a 3D3V simulation, a distribution function $f(\mathbf{x},\mathbf{v})$ has six dimensions. 
Electron and ion particle distributions tend to be Gaussian in nature as a consequence of the H-Theorem and relaxation processes present in plasmas~\cite{jackson1999electrodynamics}. If particle distribution functions are different from Gaussian distribution functions, it hints at the presence of certain resonance phenomena that act only on a small part of the plasma populations, such as Landau damping or bump-on-tail. A Gaussian centered on a given velocity, different from zero, shows the presence of a beam with that specific bulk velocity. Similarly, an increase in the standard deviation of the Gaussian reflects a rise in the local plasma temperature. For these reasons, a GMM-based compression method is called \emph{physics-aware}. This work provides the following contributions: 
\begin{itemize}
\item We develop a methodology to reduce the memory and computational demands of GMM, applying it to an intermediate histogram instead of particle data directly. 
\item We develop a GPU implementation of the histogram-GMM pipeline and execute it during spare GPU cycles in the PIC simulation to compress plasma distribution functions.
\item We show physics-aware compression using Gaussian parameters, and compare the information loss, compression rate and performance with several established compression algorithms.
\end{itemize}

\section{Background \& Related Work}
\label{sec:bggmm}
\begin{figure}[h] 
    \centering
    \includegraphics[width=0.75\linewidth]{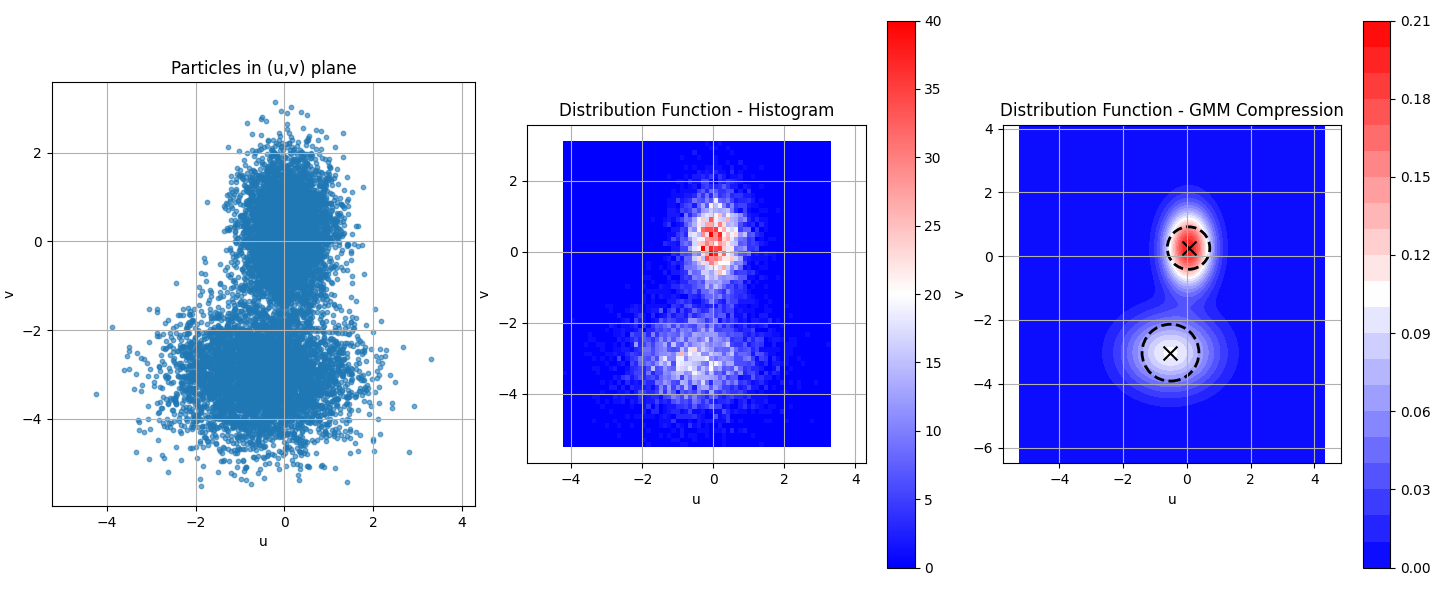} 
    \caption{The distribution function of 10,000 particles can be approximated with an $64\times64$ histogram (central panel) or a GMM approach storing the weights, the center and covariance matrices of two Gaussians (rightmost panel).}
    \label{fig:GMM2D}
\end{figure}

Large-scale PIC simulations compute the trajectories of billions of computational particles, generating Terabyte-size data, including distribution functions. Fig.~\ref{fig:GMM2D} shows different possibilities of storing particle information for particle distribution functions. On the leftmost panel, there is the raw information of particles velocity: in this plot, each dot corresponds to a particle, requiring the storage of $2 N_p$ floating point data, where $N_p$ is the number of particles. One compression scheme for particle distribution function, is to introduce a 2D grid in the space $u,v$ to discretize the distribution and function and count how many particles belong to each bin. This data reduction technique allows us to encode our distribution function in a $N_g \times N_g$ grid. The approach we pursue in this work is to use the GMM technique to fit the particle distribution function with a relatively small number of Gaussians (1-20) and store only the Gaussian parameters for the reconstruction. In the example of Fig.~\ref{fig:GMM2D}, we only need the parameters of two Gaussians (12 floating point values for this two-dimensional example) to capture the complexity of the given distribution function.

The basic idea of the GMM technique is to represent a probability distribution as a weighted sum of multiple Gaussian components, each defined by a mean vector and a covariance matrix~\cite{bishop2006pattern,ghojogh2019fitting}. Mathematically, a GMM models a probability density function (pdf) $p(\mathbf{x})$ as a weighted sum of $M$ Gaussian components:
\begin{align}
    p(\mathbf{x}) &= \sum_{i=1}^M \alpha_i \mathcal{N}(\mathbf{x} | \boldsymbol{\mu}_i, \boldsymbol{\Sigma}_i), \label{eq:gmm}
\end{align}
where $\alpha_i$ are the mixture weights, $\boldsymbol{\mu}_i$ are the means, and $\boldsymbol{\Sigma}_i$ are the covariance matrices. Given $N$ observed data, each one denoted as $\mathbf{x}_n$, the GMM parameters $\{\alpha_i, \boldsymbol{\mu}_i, \boldsymbol{\Sigma}_i\}$ are estimated using the Expectation-Maximization (EM) algorithm~\cite{mclachlant97EM_algorithm}, which maximizes the likelihood function $\mathcal{L}( \boldsymbol{\theta}|\mathbf{x})$, defined as: $\mathcal{L}( \boldsymbol{\theta}|\mathbf{x})=\sum_{i=1}^N\left[ \sum_{j=1}^M \alpha_j \mathcal{N}(\mathbf{x}_n | \boldsymbol{\mu}_j, \boldsymbol{\Sigma}_j) \right]$. 
The EM algorithm consists of a two-step iteration:
\begin{itemize}
    \item \textbf{E-step:} Compute responsibilities for each Gaussian component for each observed data:
    \begin{align}
        \gamma_{i}(\mathbf{x}_n) &= \frac{\alpha_i \mathcal{N}(\mathbf{x}_n | \boldsymbol{\mu}_i, \boldsymbol{\Sigma}_i)}{\sum_{j=1}^M \alpha_j \mathcal{N}(\mathbf{x}_n | \boldsymbol{\mu}_j, \boldsymbol{\Sigma}_j)}. \label{eq:responsibilities}
    \end{align}
    \item \textbf{M-step:} Update the parameters:
   \begin{align}
    \alpha_i^{\text{new}} &= \frac{1}{N} \sum_{n=1}^N \gamma_i(\mathbf{x}_n) \label{eq:alpha_update}
    \end{align}
    \vspace{-\baselineskip}
    \begin{align}
    \boldsymbol{\mu}_i^{\text{new}} &= \frac{\sum_{n=1}^N \gamma_i(\mathbf{x}_n) \mathbf{x}_n}{\sum_{n=1}^N \gamma_i(\mathbf{x}_n)} \label{eq:mu_update}
    \end{align}
\vspace{-\baselineskip}
    \begin{align}
    \boldsymbol{\Sigma}_i^{\text{new}} &= \frac{\sum_{n=1}^N \gamma_i(\mathbf{x}_n) (\mathbf{x}_n - \boldsymbol{\mu}_i)(\mathbf{x}_n - \boldsymbol{\mu}_i)^\top}{\sum_{n=1}^N \gamma_i(\mathbf{x}_n)} \label{eq:sigma_update}
    \end{align}
\end{itemize}
The EM-GM technique is particularly suitable for describing the probability distribution associated with physical quantities since this algorithm conserves, exactly at each iteration -- regardless of whether it has converged or not -- the moments of the observed data up to the second moment~\cite{chen2021unsupervised}. This ensures the conservation of physical bulk quantities.

The GMM technique has been used before for compressing scientific data in several research fields. For instance, it has been used to compress image data~\cite{sun2021_gmm_image} and medical ECG data~\cite{Sahoo2024_gmmECG}. In the area of computational plasma physics, GMM has been used for checkpointing and restart~\cite{chen2021unsupervised} and as a post-processing tool to characterize magnetic reconnection regions~\cite{Dupuis_2020_gmm_magneticReconnection} in PIC simulations.

Several compression techniques have been developed for scientific computing data~\cite{li2018data,di2024survey}, consisting of floating-point number. Among the most used and successful compression methods, there is MGARD (MultiGrid Adaptive Reduction of Data)~\cite{liang2021mgard+}, developed at Oak Ridge National Lab for multi-level (inspired by Multigrid methods) data error-bounded lossy compression. The basic compression mechanism is based on decomposing into multiple levels (or grids). At each level, the coarse grained components are separated from the finer details, leading to an hierarchy of coarser and finer representations of the data. Another widely used compression scheme, targeting floating-point data and HPC simulation result, is ZFP~\cite{lindstrom2025zfp}, developed at Lawrence Livermore National Laboratory. The basics mechanisms are based on dividing the input into fixed-size blocks, reordering into a Z-order curve, and applying a customized Discrete Cosine Transform-like Transform, similarly to JPEG. In addition, another popular compression scheme is SZ (Squeeze in short)~\cite{di2016fast}, developed at the Argonne National Laboratory. This method is based on a prediction step, based on a Lorenzo predictor~\cite{ibarria2003out} or linear regression, predict the value of each data point based on its neighboring values, error quantization and Huffmann encoding. A comparison of our proposed GMM-based compression with other compression schemes for scientific computing is presented in Table~\ref{tab:comparison}.

In this work, we design and implement a compression method based on GMM, and demonstrate its effectiveness in the iPIC3D code, a massively parallel PIC code~\cite{markidis2010multi}. 
The code is designed for space plasma simulations, focusing on studying phenomena such as plasma-wave interactions~\cite{yu2018pic}, magnetic reconnection~\cite{peng2015energetic}, collisionless shocks~\cite{peng2015kinetic}, and global planetary magnetosphere dynamics~\cite{peng2015formation}. iPIC3D uses the implicit moment PIC method \cite{mason1981implicit}, distinguishing it from standard explicit PIC approaches. This method relaxes numerical stability constraints related to time step and grid resolution, enabling simulations of large-scale systems such as planetary magnetospheres. iPIC3D is implemented in C++ with support for MPI and OpenMP, and it has demonstrated strong scalability, running on up to one million MPI processes~\cite{markidis2016epigram}. Additionally, it supports heterogeneous architectures, including supercomputers with AMD and NVIDIA GPUs. To compare the results of our GMM-based compression scheme and other standard compression schemes, we developed an ADIOS~2~\cite{godoy2020adios} I/O iPIC3D module to leverage different compression schemes.
\begin{table}[h]
    \centering
    % \renewcommand{\arraystretch}{1.2}
    % \small 
    \resizebox{\textwidth}{!}{
    \begin{tabular}{p{3.5cm} p{5.5cm} p{5.5cm}}
        \toprule
        \textbf{Feature} & \textbf{GMM-Based Compression} & \textbf{Other Compression Schemes (ZFP, SZ, MGARD, ...)} \\
        \midrule
        \textbf{Compression Type} & Lossy, physics-aware Gaussian approximation & Lossy or lossless, numerical-based approximations \\
        \midrule
        \textbf{Data Representation} & Stores Gaussian parameters (mean, covariance, weight) instead of full velocity distributions & Encodes data using fixed-size blocks, wavelets, or error-controlled quantization \\
        \midrule
        \textbf{Physics Preservation} & Retains key plasma physics properties (bulk velocity, temperature, beams, heating) & May distort physical properties, especially in structured plasma distributions \\
        \midrule
        \textbf{Compression Ratio} & Adaptive, based on the number of Gaussians; high reduction rates for Gaussian-like distributions & Fixed compression ratio; depends on error tolerance settings. \\
        \midrule
        \textbf{Computational Cost} & Iterative fitting; GPU-accelerated for in-situ execution & Faster for standard lossless/lossy compression; They can be computationally expensive. \\
        \midrule
        \textbf{Error Control} & Adjusts Gaussian component number dynamically to control error or prune Gaussian with relatively small weight & Provide absolute or relative error bounds \\ 
        \midrule
        \textbf{Interpretability} & Directly relates to plasma physics; Gaussian parameters can be analyzed for physical insights & Mathematical compression.\\
        \bottomrule
    \end{tabular}
    }
    \caption{Comparison of GMM-Based Compression with other scientific data compression schemes.}
    \label{tab:comparison}
\end{table}

\section{Methodology}

\subsection{Weighted Gaussian Mixture Model}
We employ the GMM on particle velocity distribution functions to compress the data and perform in-situ analysis. Since we are mainly interested in full-scale three-dimensional simulations, for the feasibility of real-time in-situ compression, we treat each simulation subdomain independently, and we pre-process the raw particle data by binning the velocities into a histogram. Specifically, from the full 6D distribution function, we derive the 3D velocity distribution function integrating over each simulation subdomain: $f(u,v,w)=\int f(\mathbf{x},\mathbf{v})\text{d}\mathbf{x}$. Then, from the 
3D $f(u,v,w)$, we create three 2D histograms that represent the reduced-dimensional distribution functions $f_1(u,v)$, $f_2(v,w)$ and $f_3(u,w)$. These histogram pdfs are eventually given as input data to GMM, thus, for each pdf we have a constant $N_b \times N_b$ number of observed data, where $N_b$ is the number of histogram bins. 

To retain the correct velocity distributions, we utilize a modified (weighted) GMM algorithm that assigns a weight to each observed data point based on the corresponding bin counts. The GMM fitting process relies on the EM algorithm, as detailed in Section~\ref{sec:bggmm}, and the weight of each input data ($w_n$) is taken into account in the E step. Thus, Eqs.~\ref{eq:alpha_update},~\ref{eq:mu_update},~\ref{eq:sigma_update} are modified to Eqs.~\ref{eq:alpha_update_weighted},~\ref{eq:mu_update_weighted},~\ref{eq:sigma_update_weighted}, respectively:
\begin{align}
\alpha_i^{\text{new}} &= \frac{ \sum_{n=1}^N w_n \gamma_i(\mathbf{x}_n)}{\sum_{n=1}^N w_n}, \label{eq:alpha_update_weighted} \\
\boldsymbol{\mu}i^{\text{new}} &= \frac{\sum{n=1}^N w_n \gamma_i(\mathbf{x}_n) \mathbf{x}_n}{\sum_{n=1}^N w_n \gamma_i(\mathbf{x}_n)}, \label{eq:mu_update_weighted} \\
\boldsymbol{\Sigma}i^{\text{new}} &= \frac{\sum_{n=1}^N w_n \gamma_i(\mathbf{x}_n) (\mathbf{x}_n - \boldsymbol{\mu}_i^{\text{new}})(\mathbf{x}_n - \boldsymbol{\mu}_i^{\text{new}})^\top}{\sum_{n=1}^N w_n \gamma_i(\mathbf{x}_n)}. \label{eq:sigma_update_weighted}
\end{align}

Our GMM implementation guarantees the generality and robustness of the algorithm. Specifically, we implement an automatic tuning of the number of Gaussians in the mixture based on the mixing weights. Every ten EM iterations, a kernel checks whether any component has a mixing weight below a certain threshold; if so, the component is pruned before proceeding with the next iteration, and the weights of the other components are rescaled. To prevent excessively reducing the number of components, Gaussians are pruned one at a time. Thus, we start the GMM with $M$ active components, and we end it with $1\leq\hat{M}\leq M$ Gaussians.
Additionally, we ensure the algorithm's numerical stability by implementing safety checks on the covariance matrix of each component. These checks guarantee that the property of being symmetric positive definite is maintained at every iteration. If, due to floating-point arithmetic, the matrix has a negative determinant, we increase the values of the main diagonal elements.

Normalization of the observed data is another beneficial step that enhances the algorithm's numerical stability. By rescaling the data to the $[-1,1]$ range, we can work with values several orders of magnitude larger than the original data, making them less sensitive to rounding errors and floating-point arithmetic limitations.

The GMM algorithm is highly sensitive to the initial parameter estimates, and a well-chosen initialization is crucial for ensuring proper convergence. When in-situ data analysis (DA) is performed for the first time in the simulation, GMM is initialized with uniform mixing coefficients, variances equal to the species temperatures, and random means to ensure that the Gaussians cover the entire data range. In the subsequent DA steps, GMM is initialized using the parameters estimated in the previous DA cycle. The more frequently DA is executed, the less the data changes between consecutive GMM runs, leading to increasingly accurate initial parameter estimates.

In our integrated histogram-GMM DA pipeline, all calculations, beginning with velocity binning, are executed on the GPU, thus minimizing host-device data transfers. Specifically, raw particle data and histogram pdfs are not copied to the host, allowing real-time integration within the simulation workflow. The results of weighted GMM include parameters like mean vectors, covariance matrix, and weights for all clusters, along with some auxiliary values such as the number of EM steps and the log-likelihood. After GMM has converged, the parameters can be copied back to the host and then written to disk in different formats, such as JSON and BP5. 

Regarding performance concerns, the buffers in the data analysis pipeline are reused in the whole life cycle of simulations, thereby minimizing latency caused by configurations. Customized reduction kernels are employed in the weighted GMM implementation for cross-platform compatibility. Tailored pre- and post-processing techniques are applied to reduce kernel launches in GMM regression.

\subsection{ADIOS 2 in iPIC3D}
In this work, we use ADIOS 2 to implement I/O of GMM-based compression and compare the performance of our GMM-based compressor with other established compressors. ADIOS 2 enables asynchronous (deferred) data output and provides operators for compression. Each particle attribute (position, velocity, charge, ID), in Structure-of-Array buffer, is mapped to an ADIOS 2 variable, then is output to Binary-Pack 5 (BP5) files.

The iPIC3D \texttt{initOutputFiles} function initializes the ADIOS 2 environment, defines output file structures, and sets up data writing parameters. Particle data is written in a structured format using the \texttt{appendParticleOutput} function, detailed in List.~\ref{list_appendParticleOutput}. This function captures the simulation cycle and appends particle properties, such as position, velocity, charge, and ID, to the output file.  It also employs ADIOS 2's \texttt{BeginStep} and \texttt{EndStep} to utilize deferred writes. 
\begin{lstlisting}[language=C++, caption={Appending Particle Output.}, label={list_appendParticleOutput}] 
void ADIOS2Manager::appendParticleOutput(int cycle) { 
    ...
   engineParticle.BeginStep();
   auto cycleVar = _variableHelper<int>(ioParticle, "cycle");
   engineParticle.Put<int>(cycleVar, cycle);
   ...
   engineParticle.EndStep();
} \end{lstlisting}

A key benefit of ADIOS 2 is its built-in data compression operation function, which enables the specification of a compressor engine and the following compression of simulation data prior to storage. For example, List.~\ref{list_dataCompressionADIOS2} details the instruction needed to apply compression to the variable \verb|var| with the SZ compressor. The ADIOS 2 interface decouples the data and the main application code from the actual compression algorithm, this allows great flexibility since we can easily change the compressor back-end without the need to restructure the I/O.
\begin{lstlisting}[language=C++, caption={Apply compression to data with a given engine.}, label={list_dataCompressionADIOS2}] 
var.AddOperation(adios2::ops::LossySZ, {{"accuracy", "0.00001"}}); 
\end{lstlisting}

In this work, we evaluate the performance and effectiveness of several non-physics-aware compression methods and compare them to our physics-aware GMM method. Specifically, we evaluate lossy methods, including SZ, ZFP, and MGARD, and lossless methods like BLOSC2 and BZIP2. All compressors are executed on the CPU and take host memory buffers as input; no GPU is involved in the process.

\subsection{Compression Performances and Information Loss Evaluation}
In this work, we analyse the computational performances and the compression rate of each compression method. Additionally, we assess the information loss in the compression stage comparing the compressed data with the ground truth.

Since in this work we focus on the particle velocity distribution, we evaluate the accuracy of the compressed data by employing the Jensen-Shannon Divergence (JSD) \cite{lin1991JSD_metric}. The JSD is an information-theoretic metric for comparing probability distributions, so it can be used to assess how well the compressed data retains the features of the original distribution. The JSD for two probability distributions $P$ and $Q$ is evaluated as reported in Eq. \ref{eq_jensenShannonDiv}:
\begin{equation}
    JSD(P||Q) = \dfrac{1}{2}D\left(P||M \right)  + \dfrac{1}{2}D\left(Q||M \right),
    \label{eq_jensenShannonDiv}
\end{equation}
where $M$ is the mixture of the distribution $M=(P+Q)/2$ and $D(\cdot)$ denotes the Kullback-Leibler divergence \cite{kullbackLeibler1951_metric}, defined as:
\begin{equation}
    D(P||Q) = \sum_{n=1}^Np(\mathbf{x}_n)\log\left( \dfrac{p(\mathbf{x}_n)}{q(\mathbf{x}_n)} \right). 
    \label{eq__kullbackLeiblerDiv}
\end{equation}

The result of JSD falls between 0 and $\ln2$. The lower the result, the closer the two distributions are to each other.

Additionally, we assess the quality of GMM, with Bayesian Information Criterion (BIC), reported in Eq. \ref{eq_bic}:
\begin{equation}
\text{BIC} = -2 \ln(\mathcal{L}) + k \ln(N),
\label{eq_bic}
\end{equation}
where $\mathcal{L}$ denotes the maximized likelihood function at the end of the EM algorithm, $k$ is the number of parameters estimated by the model and $N$ the number of observed data. In the case of multivariate GMM applied to $d$ dimensional data $k=M(1 + d(d+3)/2 )$, where $M$ is the number of Gaussians in the mixture. The BIC metric penalizes mixtures with a high number of components. 

We highlight here some crucial aspects related to ground truth identification and information loss evaluation for the GMM compression technique. In PIC simulations, due to the finite number of computational particles, the numerical distribution function is a noisy approximation of the true, smooth physical pdf, which is the correct ground truth, but is never available. Additionally, in our implementation, we utilize pre-processed binned data as input for the GMM algorithm. Therefore, we can consider the pdf extrapolated from the histogram (which is still noisy) as the ground truth. It is important to note that this pdf does not exactly match the numerical pdf since binning acts as a smoother. On the other hand, the GMM method inherently produces a smooth pdf that may ultimately offer a more accurate approximation of the true physical pdf than the noisy one provided by the computational particles. For this reason, evaluating the accuracy of the GMM pdf solely with a metric based on point-wise comparison with a noisy pdf, such as the JSD metric, may yield incomplete information.

In our analysis, we calculate the JSD metric to show the differences between the histogram and GMM pdf. We also compute the JSD between the raw particle data pdf and the histogram by binning the particle velocities into a highly refined $500\times500$ histogram (labeled as \textit{original} data in the following) and comparing it with the linearly interpolated histogram used as the first stage of the GMM compression. Since the EM algorithm inherently ensures the conservation of quantities like the mean and variance of the data, we extend the evaluation of the GMM compression accuracy with a qualitative comparison between the GMM pdf and the histogram pdf shape.

In order to asses the performance, namely processing time and compression ratio, of the five standard compression engines and our GMM-based algorithm, we deploy all of them in the same GEM simulation run to ensure the utilization of the same raw data.

\section{Experimental Setup}
\begin{figure}[h!]
    \centering
    \includegraphics[width=0.6\linewidth]{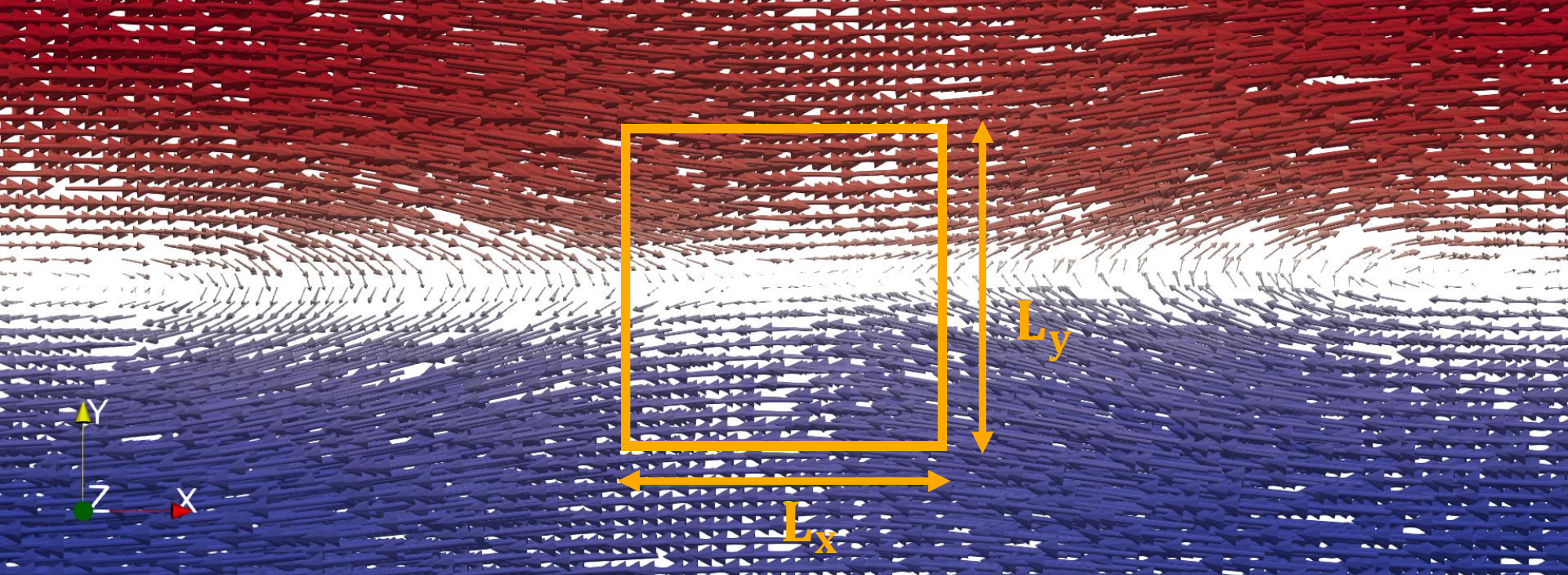}
    \caption{Magnetic field line at the x-point during magnetic reconnection. The region where we take the data, with dimensions $L_x=L_y=1.42$ $d_i$, $L_z=2$ $d_i$ is highlighted in orange. }
    \label{fig_xPointBx}
\end{figure}

As a main demonstration for the GPU accelerated GMM compression, we run a 3D version of the standard plasma physics benchmark problem, called GEM challenge, with the iPIC3D code. This test mimics the plasma condition in the Earth magnetotail, initiates magnetic reconnection phenomena perturbing the magnetic field topology along a central line in anti-parallel configuration of the magnetic field, without the presence of a background magnetic field.  The size of the simulation box, in units of ion skin depth ($d_i$), is $10\times10\times10$, the domain is discretized with a uniform Cartesian grid of $133\times133\times135$ cells, and we run the simulation for 3,000 steps. The domain is divided into $7\times7\times5$ subdomains, each one assigned to one MPI rank.
Fig.~\ref{fig_xPointBx} shows an example of magnetic field configuration during magnetic reconnection. The area in the orange box represents the region from which we obtain the particle data, corresponding to the subdomain at the center. This region has dimensions of $L_x=L_y=1.42$ $d_i$ and $L_z=2$ $d_i$.
We run GMM every 50 simulation cycles, with an initial number of 12 Gaussians, for a maximum of 100 EM cycles. The automatic pruning method removes components with a mixing coefficient lower than 0.005. Raw particle data is pre-processed by binning the velocities into fixed-range histograms of $200\times200$ bins. 

We evaluate the performance of the different compression engines with a smaller GEM simulation, with a grid of $64\times64\times64$ cells divided into $2\times2\times2$ subdomains. We run the simulation for 4000 cycles. The output frequency for particle data is set to every 500 cycles, and particle velocity and charge data are written to BP5 files using ADIOS 2. For evaluation, we fix the number of GMM components to eight and disable features like pruning, with a $100\times100$ histogram as its pre-processor. The accuracy parameter of the lossy compressors is set to $0.00001$, $0.01$, and $0.01$ for SZ, ZFP, and MGARD, respectively, to keep the output size as similar as possible. As for the lossless methods, BZIP2 uses $blockSize100k = 5$ and BLOSC2 uses $blosclz$ and $clevel = 9$.

We perform our tests on two main systems, involving both Nvidia and AMD GPUs. The first system is one node of the Sleipner cluster at KTH equipped with a Grace-Hopper 200 480GB accelerator. We use this machine to compare the GMM algorithm with different compressors in terms of achievable compression ratio and execution time. The iPIC3D production runs involving in-situ GMM-based data compression are executed on the LUMI-G supercomputer. Each LUMI-G computing node is equipped with four AMD MI250x GPUs (each one with two GCDs and 128 GB HBM memory) and a 64-cores AMD EPYC Trento with a total of 512 GB CPU memory.

\section{Results}

\subsection{GMM Compression Accuracy}
In this section, we evaluate the accuracy of the GMM compression technique by comparing the reconstructed velocity pdf with the one obtained from the histogramming process, for both electron and ion species at the end of the large magnetic reconnection simulation.
We primarily evaluate the overall GMM compression accuracy by qualitatively comparing the GMM pdf and the histogram pdf shape. Additionally, we report the point-wise difference between the two pdfs to spot regions where they might be greatly different. 

\begin{figure}[!h]
    \centering
    \includegraphics[width=0.99\linewidth]{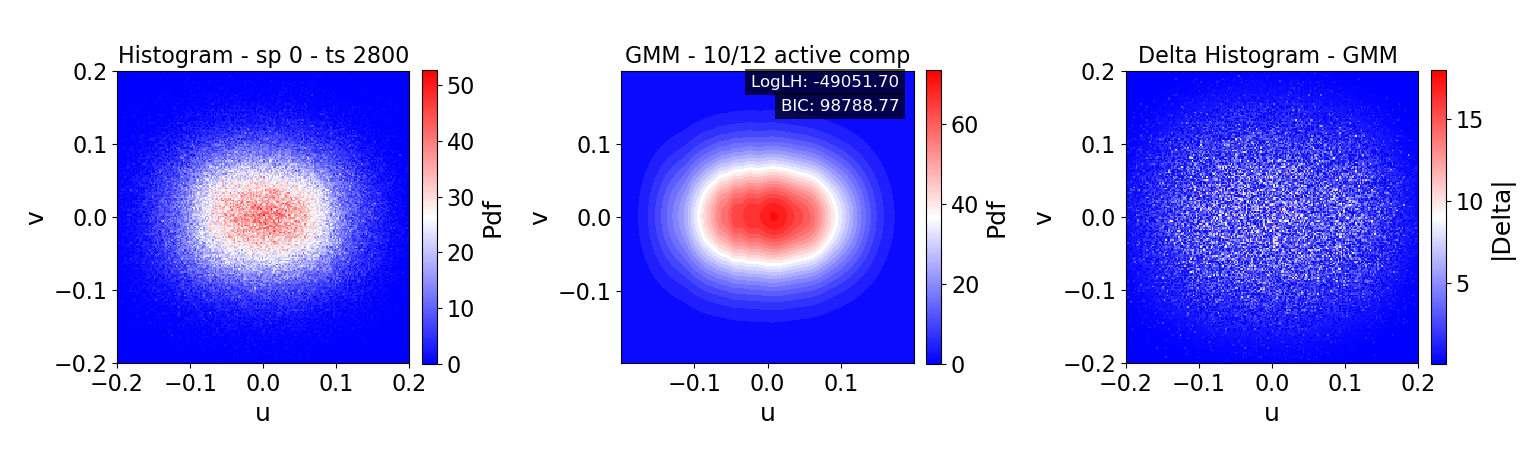}
    \caption{Electron $uv$ velocity pdf at the x-point location. Left panel: 2D pdf obtained from the histogram; central panel: 2D pdf reconstructed by GMM; right panel: absolute difference between the true and reconstructed pdf.}
    \label{fig_GMMout2DElectrons}
\end{figure}

The three panels in Fig.~\ref{fig_GMMout2DElectrons} show, respectively, from left to right, the histogram 2D pdf, the GMM reconstructed 2D pdf, and the point-wise absolute difference between the two pdfs for the electron species $uv$ velocity. The JSD metric between the histogram and GMM pdfs is 0.0157. Data are retrieved from the subdomain containing the x-point. 

\begin{figure}[!h]
    \centering
    \includegraphics[width=0.95\linewidth]{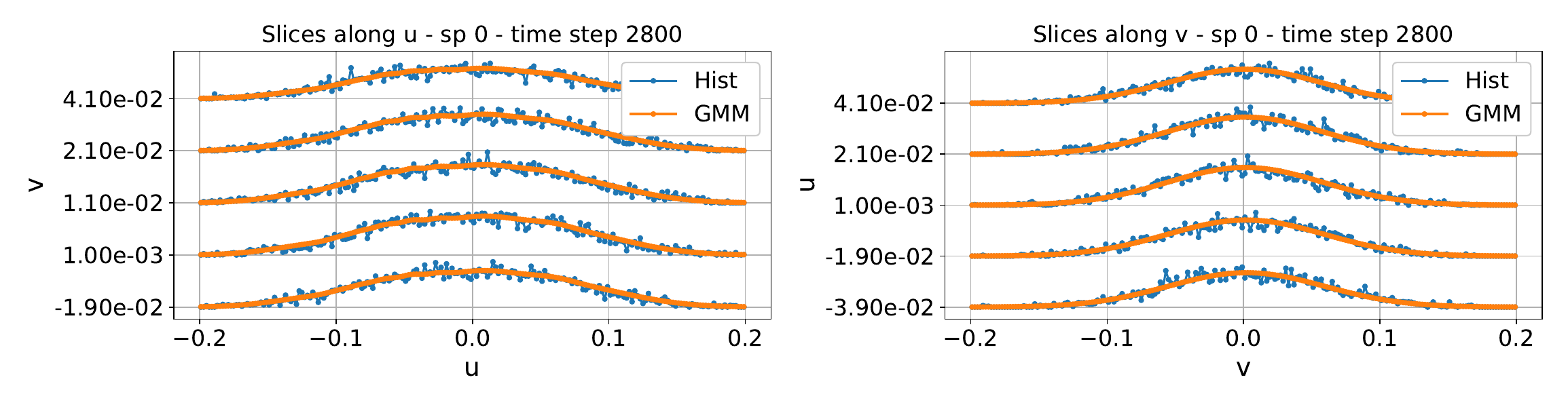}
    \caption{Electron $uv$ velocity pdf at the x-point location. 1D plots obtained as slices of the 2D pdf. In blue the true pdf obtained from the histograms, in orange the GMM reconstructed pdf.}
    \label{fig_GMMoutSlicesElectrons}
\end{figure}

To provide a better comparison between the histogram and the GMM pdfs, we report in Fig.~\ref{fig_GMMoutSlicesElectrons} 1D slices of the 2D pdf shown in Fig.~\ref{fig_GMMout2DElectrons}, along both the two velocities \textit{u} and \textit{v}. Specifically, the left panel shows the 1D profile of the pdf along the \textit{u} direction for five different \textit{v} velocities, while the right panel reports the 1D profile of the pdf along the \textit{v} direction for five \textit{u} velocities. The GMM reconstructed pdf accurately reproduces the shape of the original pdf.

\begin{figure}[h!]
    \centering
    \includegraphics[width=0.99\linewidth]{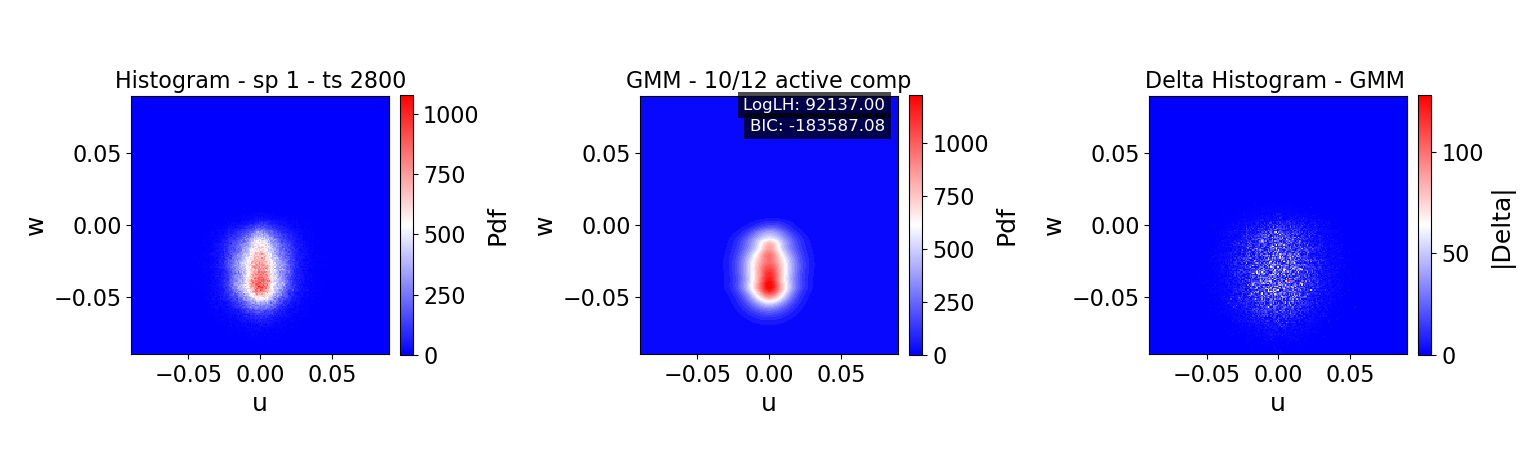}
    \caption{Ion $uw$ velocity pdf at the x-point location. Left panel: 2D pdf obtained from the histogram; central panel: 2D pdf reconstructed by GMM; right panel: absolute difference between the true and reconstructed pdf.}
    \label{fig_GMMout2DIons}
\end{figure}

The three panels in Fig.~\ref{fig_GMMout2DIons} show respectively, from the left to the right, the histogram 2D pdf, the GMM reconstructed 2D pdf, and the point-wise absolute difference between the two pdfs for the ion species $uw$ velocity. The JSD metric between the histogram and GMM pdfs is 0.0049. Data are retrieved from the subdomain containing the x-point.

\begin{figure}[h!]
    \centering
    \includegraphics[width=0.95\linewidth]{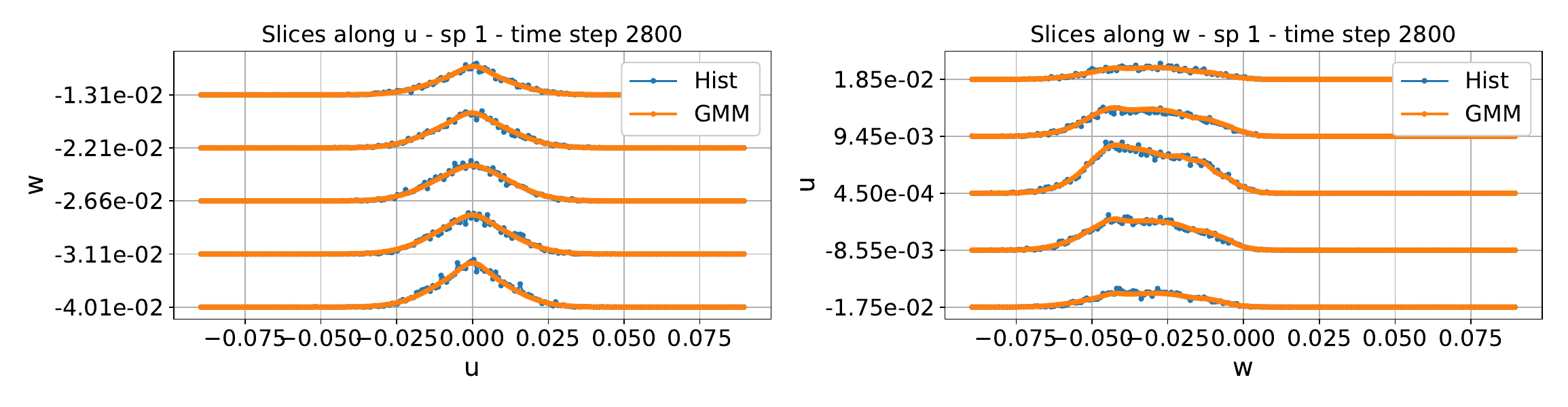}
    \caption{Ion $uw$ velocity pdf at the x-point location. 1D plots obtained as slices of the 2D pdf. In blue the true pdf obtained from the histograms, in orange the GMM reconstructed pdf.}
    \label{fig_GMMoutSlicesIons}
\end{figure}

In Fig.~\ref{fig_GMMoutSlicesIons} 1D slices of the 2D pdf shown in Fig.~\ref{fig_GMMout2DIons}, along both the two velocities \textit{u} and \textit{w}, are reported. Specifically, the left panel shows the 1D profile of the pdf along the \textit{u} direction for five different \textit{w} velocities, while the right panel reports the 1D profile of the pdf along the \textit{w} direction for five different \textit{u} velocities. As for the electron pdf, GMM is capable of accurately reproducing the shape of the original pdf.

\subsection{Compression Algorithms Evaluation}
\begin{figure}[h!]
    \centering
    \includegraphics[width=0.90\linewidth]{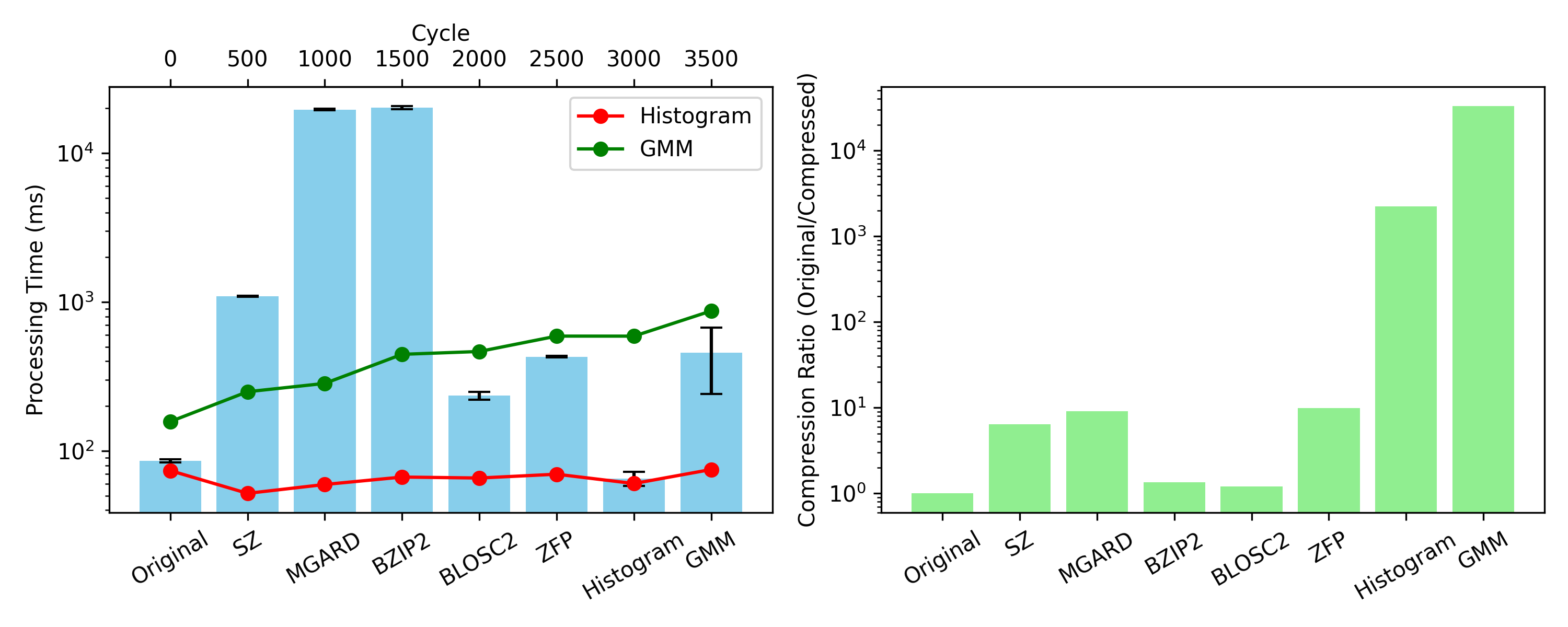}
    \caption{Left panel: processing time (compression and I/O); right panel: compression ratio for different compression methods. The GMM time includes the time required to pre-process data with histogram. The red and green lines are the histogram and GMM processing time in different simulation cycles.}
    \label{fig:compressionRateandTime}
\end{figure}

We detail in this section the performance of the different compression methods tested in this work on the small GEM simulation.
Fig.~\ref{fig:compressionRateandTime} shows the required time for compression and the compression ratio achieved by different algorithms. The histogram-GMM compressor requires a processing time in the order of 100s ms, which is comparable to other engines like BLOSC2, SZ, ZFP and lower than engines like MGARD and BZIP2. The GMM with eight components reaches a compression ratio of 14 when compared to histogram data, and up to 10,000 when compared to raw particle data. GMM compression time is growing as the simulation goes, due to the increment in the number of EM iterations required to converge. This behavior is expected since the velocity pdf complexity increases as the simulation evolves. However, the increment in time can be limited by tuning the maximum number of EM iterations.  

\begin{figure}[h!]
    \centering
    \subfloat[]{\includegraphics[width=0.48\textwidth]{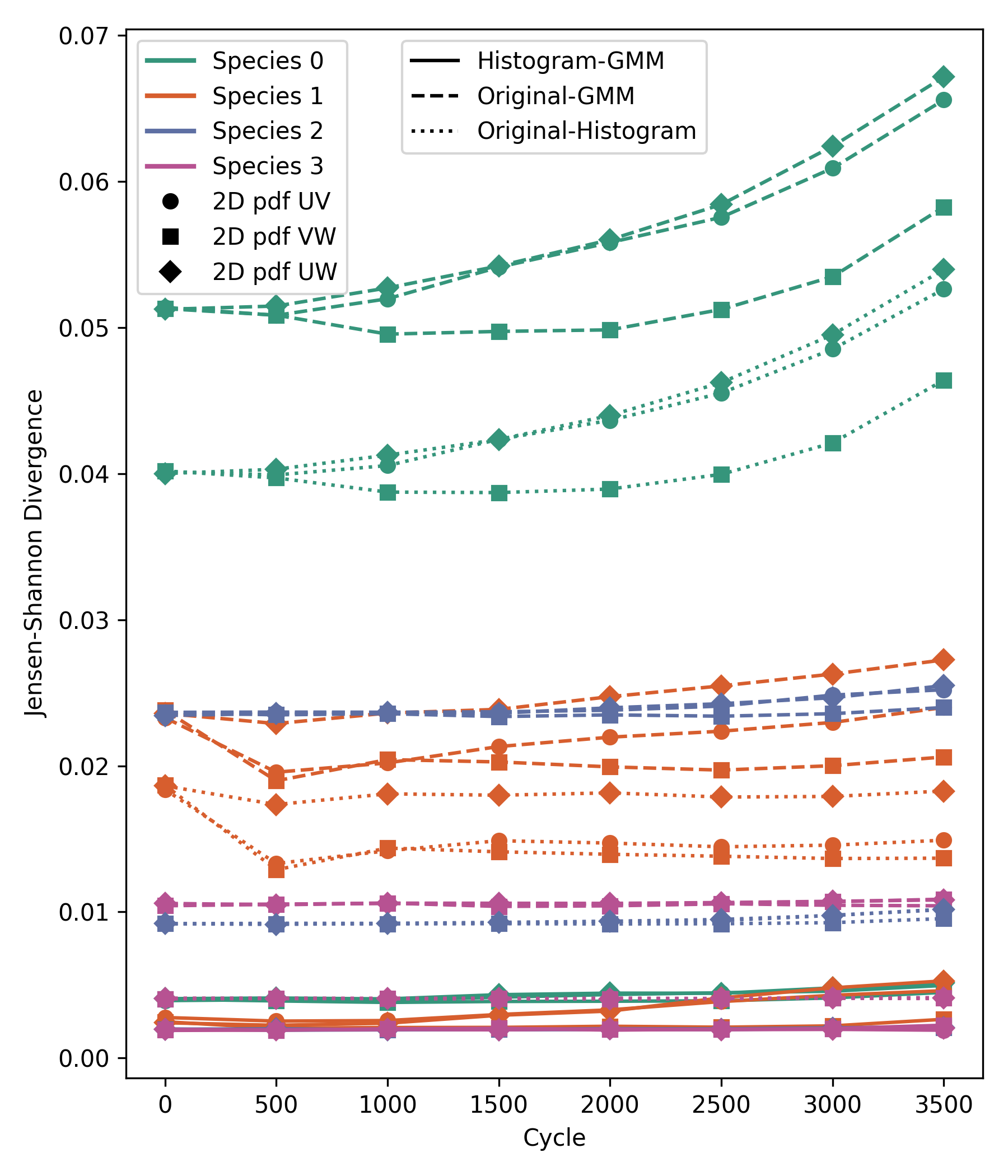}}
    \hspace{0.01\textwidth}
    \subfloat[]{\includegraphics[width=0.48\textwidth]{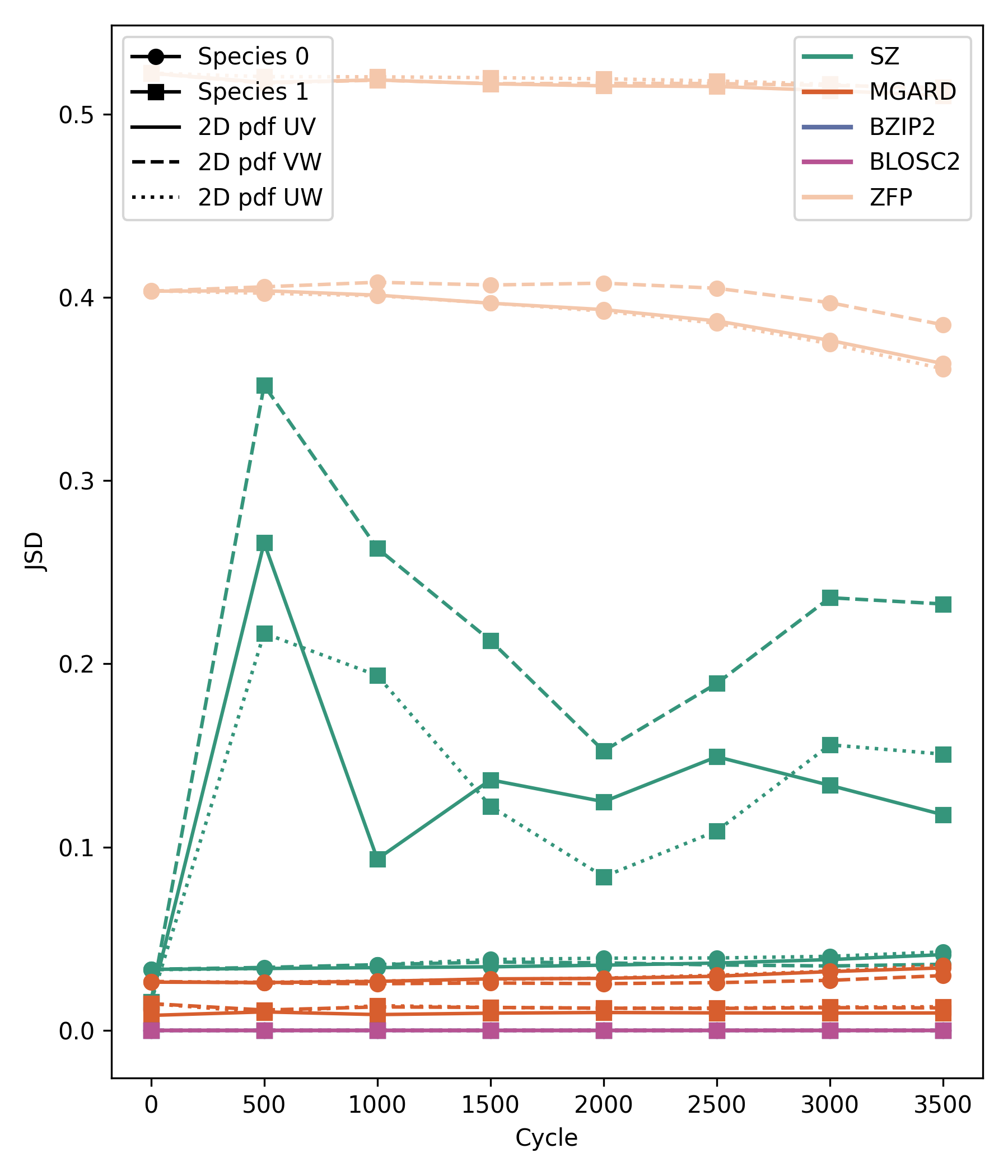}}
    \caption{(a): JSD between original data, histogram and GMM for different species and different 2D velocities pdf in subdomain 0. 
    (b): JSD between compressors and the original particle data of subdomain 0; 
    BZIP2 and BLOSC2 has zero JSD since they are lossless compressors. In both panels, species 0 (1) are main electrons (ions), while species 2 (3) are background electrons (ions).}
    \label{fig:JSD}
\end{figure}

In our specific test case, the velocity pdfs do not significantly differ from Gaussians, thus both the histogram and GMM achieve a much higher compression ratio than non-physics-aware methods, with lower information loss (JSD < 0.1), as shown in Fig.~\ref{fig:JSD}. Clearly, the JSD metric is zero for lossless compressors like BZIP2 and BLOSC2.

\section{Discussion \& Conclusion}
This work presented a GPU-accelerated Weighted Gaussian Mixture Model approach for compressing and analyzing high-dimensional data generated by PIC simulations. Weighted GMM was integrated into the iPIC3D code, and a new I/O module using the ADIOS 2 framework was implemented to compress particle velocity distributions. We showed only a small performance overhead while still achieving significant storage reduction and maintaining key statistical properties of the data. Our results demonstrated a compression rate exceeding 100x compared to established compression methods and a lower information loss in particle velocity distribution. We overlapped the iPIC3D electromagnetic solver on the CPU with GMM calculations on the GPU, showing utilization of GPU idle cycles for real-time GMM with minimized performance overhead.  

Tuning of the GMM compressor parameters was the major challenge. Specifically, the choice of the number of GMM components critically influenced the accuracy and convergence steps since not all the distributions in the GEM simulation required the same number of clusters for adequate representation. To address this, we developed an adaptive algorithm that prunes Gaussians with negligible weights. Parameter initialization also proved difficult, particularly for complex particle distributions. Tackling this issue requires enhanced parameter initialization methods and feedback mechanisms between GMM results and subsequent analysis.

Future work will involve developing an efficient detection algorithm that, by exploiting the data compression achieved with GMM, can highlight deviations in the particle distributions from the initial pdfs in real-time while the simulation is evolving. This tool would be particularly beneficial for scientists, as it would allow them to identify relevant physical phenomena, such as heating or beam formation, with minimal computational cost.

\section*{Acknowledgment}
The authors wish to thank Måns I. Andersson for the precious discussions.
This work is funded by the European Union. This work has received funding from the European High Performance Computing Joint Undertaking (JU) and Sweden, Finland, Germany, Greece, France, Slovenia, Spain, and the Czech Republic under grant agreement No. 101093261, Plasma-PEPSC, \url{https://plasma-pepsc.eu/}.

\bibliographystyle{splncs04}
\bibliography{ICCS_GMM_paper}

\end{document}